\documentclass[preprint, preprintnumbers,showpacs, showkeys, aps, prd, amsmath, amsfonts, amssymb,
   eqsecnum, nofootinbib, floatfix, secnumarabic]{revtex4}

\usepackage{epsf}
\usepackage{epsfig}
\usepackage{graphics}
\usepackage{graphicx,bbm}
\usepackage{amssymb}
\usepackage{color}

             \newcommand{\gsim}{\lower.7ex\hbox{$\;\stackrel{\textstyle>}{\sim}\;$}}
\newcommand{\lsim}{\lower.7ex\hbox{$\;\stackrel{\textstyle<}{\sim}\;$}}

\begin{document}
\preprint{CUMQ/HEP 185}

\title{Light Neutralino Dark Matter in $U(1)^\prime$ models}

\author{Mariana Frank$^1$\footnote{Email: mariana.frank@concordia.ca}}
\author{Subhadeep Mondal$^2$\footnote{Email: tpsm2@iacs.res.in}}

\affiliation{ $^1 $Department of Physics,  
Concordia University, 7141 Sherbrooke St. West ,
Montreal, Quebec, Canada H4B 1R6,}
\affiliation{$^2$Department of Theoretical Physics, 
Indian Association for the Cultivation of Science, 
2A \& 2B Raja S.C. Mullick Road, Kolkata 700032, India}

\date{\today}

\begin{abstract}
We analyze the prospects for light neutralino dark matter in the minimal supersymmetric model 
extended by a $U(1)$ gauge group. We allow the neutralino to be an arbitrary  admixture of singlet 
and doublet higgsinos, as well as of the three gauginos, and we require agreement with the data from 
the direct and indirect dark matter detection experiments, while maintaining consistency of the 
model with the  relic density, and with the recent Higgs data from the LHC. The constraints have 
implications for the structure of the lightest neutralino as a dark matter candidate, indicating 
that it is largely singlino, and its mass can be as light as $\sim 20 $ GeV. 
\end{abstract}
\maketitle
\section{Introduction}
\label{sec:intro}
The recent discovery of the Higgs boson at the LHC \cite{Chatrchyan:2012ufa, Aad:2012tfa} indicates that 
the observed particle is consistent with the only missing link in the Standard Model (SM). At the same time, given that the 
SM is incomplete from a fundamental theoretical point of view, 
various groups have used the available data to set limits on the beyond the SM scenarios. New analyses of alternative scenarios 
must take into account the best non-refutable signal of physics beyond the SM available, the existence of 
dark matter, which is well established  by various cosmological and astrophysical observations trying to 
explain the dynamics of galaxy clusters and rotations. Dark matter (DM) is thought to form about $26\%$ of 
the mass-energy of the Universe, with the relic density  measured by WMAP \cite{Komatsu:2010fb} and 
PLANCK \cite{Ade:2013zuv} experiments, where the current value of $\Omega_{\rm DM}h^2$ is
\begin{equation}
\Omega_{\rm DM}h^2=0.1126 \pm 0.0036.
\end{equation} 
Various direct and indirect detection experiments have reported signals consistent with DM particle interpretation. 
 
The so-called direct detection experiments are underground experiments who look for the direct evidence of 
dark matter through its elastic scattering with nuclei of different target materials. Galactic weakly interacting massive particles (WIMPs) can 
collide with nuclei transferring kinetic energy to a single nucleus. Low DM masses, around 10 GeV or so are 
favoured by DAMA/LIBRA \cite{Bernabei:2008yi}, CoGeNT \cite{Aalseth:2010vx} and recently by CDMS-II 
\cite{Agnese:2013rvf} experiment, while the medium mass region of 25-60 GeV is preferred by CRESST-II 
\cite{Angloher:2011uu}. All those events lie in the regions excluded by the XENON10 and XENON100 experiments, 
which set strong limits on the DM-nucleon scattering cross-section \cite{Aprile:2012nq}. 
The prediction for the spin-independent scattering cross-section with nuclei is affected  by 
uncertainties in light quark masses and hadronic matrix elements, and is heavily dependent on assumptions about the local density of dark
matter in our galaxy.  Taking these into account, the scattering 
cross-section ranges between roughly between $10^{-44} - 10^{-46}$ cm$^2$. The most stringent constraint is 
by the LUX experiment \cite{Akerib:2013tjd}, which finds cross sections of about $10^{-45}$ cm$^2$ for a light 
dark matter candidate in the mass range 10 GeV$ \lsim m \lsim 30$ GeV.  

Complimentary to these, indirect detection experiments test cosmic rays signals in the hope of revealing 
annihilations or decays of DM particles in these fluxes, and to establish additional properties of dark matter. 
There has been a flurry of results in indirect detection experiments pointing to an excess of electrons 
and positrons. First, data from the PAMELA satellite \cite{Adriani:2008zr} showed a steep increase in the 
energy spectrum of positrons above 10 GeV and up to 100 GeV, but no excess in the antiproton versus proton 
energy spectrum compared to the background. Several groups analyzing data from the Fermi Gamma-Ray Space 
Telescope have reported the detection of a gamma-ray signal from the inner few degrees around the galactic 
center, with a spectrum and angular distribution compatible with that anticipated from annihilating dark 
matter particles. The Fermi LAT signal \cite{FermiLAT:2011ab,Ackermann:2011wa} is very well fit by a 31-40 
GeV dark matter particle annihilating into $b \bar b$, with an annihilation cross section of 
$\langle \sigma v \rangle = (1.4- 2.0) \times 10^{-26}$ cm$^3/s$. Furthermore, Fermi LAT finds that the 
angular distribution of the excess is approximately spherically symmetric and centered around the dynamical 
center of the Milky Way, showing no sign of elongation along or perpendicular to the galactic plane, and thus 
disfavoring the idea that the emission comes from milli-second pulsars. And recently,  a 3.5 keV X-ray line was detected in XMM-Newton data, 
seen in M31 and Perseus galaxy cluster \cite{Boyarsky:2014jta},  and in stacked
spectra of 73 galaxy clusters \cite{Bulbul:2014sua},  and thus appears to be a  very light DM candidate.  
 
So far, the most compelling candidate for dark matter is provided by supersymmetry (SUSY). To avoid copious decays 
of supersymmetric particles into ordinary particles, one imposes a discrete symmetry, $R$-parity, 
$R=(-1)^{3B+L+2s}$, where $B$ is the baryon number, $L$ is the lepton number, and $s$ is spin. Conservation 
of $R$-parity implies the existence of a stable lightest supersymmetric particle (LSP), which does not have 
strong or electromagnetic interactions and becomes the candidate for  
weakly interacting massive particles (WIMPs) which are thought to form dark matter. In the minimal supersymmetric model (MSSM), that role is played by the lightest 
neutralino,  assumed to be  be mainly bino. Unfortunately a bino-like neutralino is typically overproduced 
in thermal processes in the early universe so that an efficient annihilation mechanism is needed in order to 
reproduce the observed relic density. Several studies have shown that light neutralino dark matter can be 
compatible with collider data, provided one allows for non-universal gaugino masses \cite{Hooper:2002nq, 
Bottino:2002ry,Bottino:2003iu,Choudhury:2012tc}. While as of now there is no agreement on how to explain some of the 
contradictory results in direct and indirect detection experiments, the question of whether it is possible to generate a light DM 
candidate has endured. Some scenarios have abandoned supersymmetry and added {\it ad-hoc} hidden dark matter sectors 
to the SM. In these models DM can be as light as one wishes. At the same time, many studies of MSSM \cite{MSSM} and 
next-to-MSSM (NMSSM) \cite{NMSSM}  investigated if 
the these models could be compatible with the existence of a dark matter particle satisfying the experimental constraints.
 
Collider measurements of the Higgs mass, decay widths and branching ratios at the LHC have 
put strong constraints on other decays of the Higgs boson, in particular the invisible one 
\cite{CMS:yva,ATLAS:2013sla}, {\it i.e.}, into light DM particles. As the data stands at present, 
the boson found at the LHC is compatible with the SM Higgs boson, but other scenarios are not ruled out. 
The LHC measurements are thus complimentary to DM detection experiments in determining the nature and 
properties of DM.    
 
In this work we investigate the possibility that the underlying symmetry is a minimal gauge extension of 
MSSM by an additional $U(1)$ group and whether  this scenario  can provide a {\it light}~ DM candidate,  
chosen to be the LSP neutralino. This neutralino can be an arbitrary admixture of singlino, bino, 
zinos and higgsinos. We impose constraints coming from relic density, direct and indirect DM detection experiments, 
as well as the Higgs couplings at the LHC, in particular from limits on the invisible decays of the Higgs boson, to constrain 
the parameter space of the model and find the optimal admixture of neutralino components consistent with all experimental 
limits. 
 
Extending the symmetry of the MSSM by an extra $U(1)$ factor is best motivated from string-inspired 
models \cite{Cvetic:1995rj}, from breaking of some supersymmetric grand unified theory (SUSYGUTs) to the SM, or as solutions of the $\mu$ problem 
\cite{Cvetic:1997ky,muprob}. In SUSYGUT symmetries, it seems difficult to break most scenarios directly to 
$SU(2)_L \times U(1)_Y$, as most models such as $SU(5), ~SO(10)$, or $E_6$ involve an additional $U(1)$ 
group in the breaking. In  models with extra $U(1)$ gauge symmetries, known as U(1)$^\prime$ models \cite{Demir:2005ti}, the number of 
Higgs bosons is increased by an additional singlet field (S) over that of the MSSM, and the vacuum expectation 
value (VEV) of the singlet $\langle S \rangle$ is responsible generating the $\mu$ term, which allows 
Higgs fields to couple to each other \cite{Cvetic:1997ky,muprob}. The $U(1)^\prime$ charges for matter and Higgs 
doublet multiplets are not unique, but bilinear Higgs coupling terms are forbidden and replaced by a trilinear scalar coupling terms. 
The NMSSM trilinear singlet self-coupling 
term in the superpotential is forbidden  as well, and replaced by $U(1)^\prime$ $D-$terms. The Peccei-Quinn symmetry is embedded in the gauged $U(1)^\prime $ 
symmetry and the would-be axion is eaten by the new $Z^\prime$ gauge boson. An additional advantage of $U(1)^\prime$ 
models is the absence of the NMSSM domain wall problems \cite{Ellis:1986mq}, because the discrete 
$\mathbbm{Z}_3$-symmetry is embedded in the continuous $U(1)^\prime$ symmetry.

$U(1)^\prime$ extended models have important cosmological implications, since the extra states can 
modify the nature of the LSP. For instance $Z^\prime$-mediated neutralino annihilation can provide 
important, and sometimes dominant contributions to DM annihilation yielding the correct relic density 
\cite{deCarlos:1997yv}. Various studies have shown that compatibility with  relic density bounds can be 
achieved for the case of a singlino-like LSP with a small higgsino component, through the $Z^\prime$ channel annihilation \cite{Choi:2006fz}. Annihilation through a $Z$ resonance is also facilitated by enhanced couplings 
to the $Z$  \cite{Barger:2004bz}. The prospects for observation of the LSP in dark matter direct
detection experiments have been considered in \cite{Barger:2007nv}. Finally, in $U(1)^\prime$ models 
 scalar right-handed sneutrinos can also be viable dark matter candidates
due to the possibility of annihilation through the $Z^\prime$ \cite{Demir:2009kc}.


Our work is organized as follows. In Sec. \ref{sec:model} we briefly describe the salient features of the 
$U(1)^\prime$ model with a particular emphasis on the neutralino sector in \ref{subsec:neutralinos}. We then 
proceed with the numerical investigation in Sec. \ref{sec:numerical},  where we discuss the model from the viewpoint 
of DM direct and indirect and collider experiments, in Sec. \ref{subsubsec:relic} and Sec. \ref{subsubsec:annihilation}. We highlight the composition of the LSP required to satisfy these constraints in Sec. \ref{subsubsec:composition}, and the impact on the light chargino and light Higgs masses in Sec. \ref{subsubsec:chargino}. We show that the chosen parameter space  satisfies constraints   from Higgs boson  data at LHC in Sec. \ref{subsubsec:LHC}. As an example of our allowed parameter space, we provide some benchmark points for 
our investigations in Sec. \ref{subsec:BPs}. We summarize our findings and conclude in Sec. \ref{sec:conclusion}.
\section{$U(1)^\prime$ model}
\label{sec:model}
We review briefly the $U(1)^\prime$ model, with particular emphasis on the neutralino sector as relevant for the DM study. 
The superpotential for the effective U(1)$^\prime$ model is
\begin{equation}
\label{superpot} W=\lambda \widehat{S}
\widehat{H}_{u}\cdot\widehat{H}_{d} + Y_t
\widehat{U}^c\widehat{Q}\cdot\widehat{H}_u+ Y_b
\widehat{D}^c\widehat{Q}\cdot\widehat{H}_d,
\end{equation}
where we assumed that all Yukawa couplings except for $Y_t$ and $Y_b$ are negligible. 
As can be seen from (\ref{superpot}), the $\mu \hat H_u \hat H_d$ term in MSSM is replaced by the $\lambda \hat S \hat H_u \hat H_d$ term. Thus the $\mu$ parameter is generated dynamically, with a singlet scalar (S) and a Yukawa 
coupling ($\lambda $), which  resolves the $\mu$ problem of the MSSM \cite{muprob}, and  $\mu_{eff}$  is expected to be of order of the weak scale. The singlet VEV $\langle S \rangle$ also 
generates the mass of the new $Z^\prime$ boson\footnote{In a realistic $U(1)^\prime $ models, three 
additional singlets are required to both generate a sufficiently large $Z-Z^\prime$ mass splitting, 
a not-too large effective $\mu$ term (to avoid fine-tuning), and to cancel the mixed anomalies between 
the SM and $U(1)^\prime$ symmetry groups. As we wish to keep our model minimal and simple, we do not include these 
additional scalars here. Consistency with $Z^\prime$ mass measurements can be achieved by considering 
the extra singlet states to be very heavy, effectively decoupled from the (scalar and fermion) spectrum.}. Note also that, unlike in NMSSM, the singlet self-coupling term $\kappa \widehat S^3$ is disallowed.

In addition to the superpotential, the Lagrangian includes $D$-terms and $F$-terms as well as soft-supersymmetry breaking terms. These  contain additional terms with respect to the MSSM, coming from
gaugino masses $M_a$ ($a=1,1',2,3$) and trilinear couplings $A_S$, $A_t$ and $A_b$ as given below
\begin{eqnarray}
\label{vd}
&&V_D=\frac{g^2}{8}(|H_u|^2-|H_d|^2)^2+
   \frac{g_{2}^2}{2}(|H_u|^2|H_d|^2-|H_u \cdot H_d|^2)+
\frac{g_{Y'}^2}{2}({\cal Q}_{H_u}|H_u|^2+{\cal Q}_{H_d}|H_d|^2+
   {\cal Q}_S|S|^2)^2, \nonumber \\
\label{vf}
&&V_F= |\lambda |^2\left[ |H_u\cdot H_d|^2+ |S|^2 (
|H_u|^2+|H_d|^2)\right], \nonumber \\
&&V_{soft}= (\sum_{a}M_a\lambda_a\lambda_a+
   A_S\lambda S H_{u}\cdot H_{d}+
A_t Y_t \widetilde{U}^c \widetilde{Q}\cdot H_u+A_b Y_b \widetilde{D}^c \widetilde{Q}\cdot H_d+h.c.)\nonumber \\
&&~~ +m_{H_u}^{2}|H_{u}|^2 +
m_{H_d}^{2}|H_{d}|^2+m_{S}^{2}|S|^2+m_{\widetilde{Q}}^2|\widetilde{Q}|^2+
m_{\widetilde{U}}^2|\widetilde{U}|^2 +
m_{\widetilde{D}}^2|\widetilde{D}|^2+
m_{\widetilde{E}}^2|\widetilde{E}|^2+m_{\widetilde{L}}^2|\widetilde{L}|^2\, ,
\end{eqnarray}
with the coupling constant $g^{2}=g_2^{2}+g_Y^{2}$. Note the presence of the new terms proportional to $g_{Y'}^2$ 
in the $D$-terms, and ${\cal Q}_i$, the particle charges with respect to the $U(1)^\prime$ gauge group, which must obey ${\cal Q}_S+ {\cal Q}_{H_u}+{\cal Q}_{H_d}=0$ 
to allow for the existence of the term $\lambda \widehat{S} \widehat{H}_{u}\widehat{H}_{d}$ in the superpotential. 
After symmetry breaking, 
the Higgs spectrum consists of a pair of charged Higgs bosons, three neutral CP-even scalars and one pseudoscalar. 
In $U(1)^\prime $ models, the allowed Higgs mass ranges have to be reconsidered due to the additional quartic Higgs 
self-couplings from the $D$-terms. The masses obtained are generally larger, LEP bounds on
Higgs bosons with SM-like couplings to the $Z$ boson are easier to satisfy, and radiative corrections from 
top-stop loops as well as constraints from gauge coupling unification in different configurations for the 
Higgs spectrum can exist, and can satisfy the LHC bounds on the SM-like Higgs boson \cite{Frank:2012ne}. 
In what follows, we are concerned with the Higgs and neutralino sectors. We choose to vary the parameters 
within a region of phenomenological interest, selected not conflict with experimental bounds. We diagonalize 
numerically the Higgs mass matrix and impose the condition that the Higgs boson  at 125 GeV is SM-like. 
We now turn our discussion to the  neutralino sector. 
\subsection{The Neutralino Mass Matrix in $U(1)^\prime$}
\label{subsec:neutralinos}
While the chargino mass matrix depends on $U(1)^{\prime}$
breaking scale through the $\mu \to \mu_{eff}$ parameter in the mass matrix, it is formally unchanged from the MSSM.  
The neutralino sector of the $U(1)^\prime$ however is enlarged from that of the MSSM by one higgsino and one gaugino state, namely
$\tilde S$ (referred to as singlino) and ${\tilde B}^{\prime}$, with a soft SUSY breaking mass, as well as possible 
$\tilde B-\tilde B^\prime$ mixing terms. We call the  the bare state of the $U(1)^\prime$ gauge fermion $\tilde B^\prime$, 
reserving the  ${\tilde Z}^\prime$ for  the physical mixed state. The mass matrix for the six neutralinos in the
$\tilde \psi^0=({\tilde B}, {\tilde W}^3, {\tilde H}^0_d, {\tilde H}^0_u, {\tilde S}, {\tilde B}^{\prime})$ basis is given 
by a complex symmetric matrix \cite{Frank:2012ne}:
\begin{eqnarray} 
{\cal M}_{\psi^0}=\left( \begin{array}{c c c c c c c c }
M_1 &   0 & -M_Z c_{\beta} s_{W}  & M_Z s_{\beta} s_{W}  & 0  & M_K \\
0   & M_2 & M_Z c_{\beta} c_{W} & -M_Z s_{\beta} c_{W} & 0 &  0 \\
-M_Z c_{\beta} s_{W} & M_Z c_{\beta} c_{W} & 0 & -\mu_{eff} & -\mu_{\lambda} s_{\beta}& {\cal Q}_{H_d} M_{v} c_{\beta}\\
M_Z s_{\beta} s_{W} & -M_Z s_{\beta} c_{W} & -\mu_{eff} & 0 & -\mu_{\lambda} c_{\beta} & {\cal Q}_{H_u} M_{v} s_{\beta}\\
 0 & 0 & -\mu_{\lambda} s_{\beta} & -\mu_{\lambda} c_{\beta} & 0 & {\cal Q}_S M_S\\
M_K & 0 & {\cal Q}_{H_d} M_{v} c_{\beta} & {\cal Q}_{H_u} M_{v} s_{\beta} & {\cal Q}_S M_S & M_1^{\prime}\\
\end{array}\right),
 \nonumber \\
\end{eqnarray}
with gaugino mass parameters $M_1$, $M_2$, $M_1^{\prime}$,  for $U(1)_Y$, $SU(2)_L$ and $U(1)^\prime$, respectively, 
and $M_K$ for $\tilde B - {\tilde B}^{\prime}$ mixing, and where $c(s)_W \equiv \cos(\sin) \theta_W$, $\theta_W$ denotes 
the electroweak mixing angle, and $c_{\beta} \equiv \cos \beta$, with $\tan \beta=v_u/v_d$. After electroweak breaking 
there are two additional mixing parameters: $M_{v} = g_{Y^{\prime}} v$  ($v=\sqrt{v_u^2 + v_d^2}$)
and $M_S = g_{Y^{\prime}}v_S$. Moreover, the doublet-doublet higgsino and doublet-singlet higgsino mixing mass mixings 
are generated to be 
\begin{eqnarray}
\mu_{eff} = \lambda \frac{v_S}{\sqrt{2}}e^{i \theta_s} ~~~~~~~,~~~~~~~ \mu_{\lambda}
= \lambda \frac{v}{\sqrt{2}}~~.
\end{eqnarray}
The neutralino eigenstates are Majorana spinors, and the physical states 
$\tilde \chi^0_i$ are defined as $\tilde \chi^0_i= {\cal N}_{ij}\tilde  \psi^0_j$. These can be obtained  by diagonalization, 
$ {\cal N}^{\dagger} {\cal M}_{\tilde \psi^0} {\cal N}={\rm  diag}(m_{\tilde \chi_1^0} ,..., m_{\tilde \chi_6^0})$. The additional 
neutralino mass eigenstates due to new higgsino and gaugino fields encode the effects of $U(1)^{\prime}$ models.
 
Now we proceed to investigate the possibility that there exists a light DM candidate consistent with the $U(1)^\prime$ LSP neutralino.
 
\section{Numerical Analysis}
\label{sec:numerical}
We explore the parameters of the $U(1)^\prime$ model in order to pinpoint allowed masses and compositions 
of the lightest neutralino consistent with dark matter relic density constraints, as well as astrophysical 
and collider requirements. Several constraints coming from collider and various DM experiments need to be 
taken into account. We discuss this in turn. 

For the relic density, we consider the best-fit value  ($\Omega_{CDM}h^2$) provided by the PLANCK experiment 
\cite{Ade:2013zuv}, $0.1199\pm 0.0027$, and stay within $3\sigma$ experimental error bars. DM-nucleon elastic scattering 
cross-section ($\sigma_{SI}$) is constrained by XENON100 \cite{Aprile:2012nq} and LUX \cite{Akerib:2013tjd} 
experiments, with LUX providing the more stringent bound. Hence we take into account the LUX bound for our present study, 
and also consider the ATLAS measurement. 

The superparticle mass limits arising from latest LHC data are taken into account \cite{atlas:sparticle,
cms:sparticle}. The lighter chargino mass is restricted by the LEP experiments to $m_{\tilde \chi^\pm_1}>103.5$ GeV 
\cite{Beringer:1900zz}. Although this limit is weaker if the mass splitting between the lighter chargino and 
the LSP is small, we respect this limit as our goal is to look for a light neutralino DM. 

Measurements of ${\rm BR}(b\rightarrow s\gamma$) and ${\rm BR}(B_s\rightarrow \mu^+ \mu^-)$ agree quite well 
with their predicted values within the framework of the SM \cite{Lees:2012ym,Aaij:2013aka,Chatrchyan:2013bka,Aaij:2012ac}. 
Hence these branching ratios constrain the $U(1)^\prime$ parameter space. Dominant contributions to 
${\rm BR}(b\rightarrow s\gamma$) come from charged Higgs and chargino exchange diagrams. In our scenario, 
the charged Higgs is kept in the TeV range but the lighter chargino mass is slightly above the LEP limit,  
imposing a serious constraint on $b \rightarrow s \gamma$. On the other hand, ${\rm BR}(B_s\rightarrow \mu^+ \mu^-)$ 
is proportional to the 6$^{\rm th}$ power of $\tan\beta$ and inversely proportional to the 4$^{\rm th}$ power of pseudoscalar mass. 
We satisfy this constraint because, for the present study, we kept the pseudoscalar mass above TeV range as well. 
 
One of the three CP-even Higgs bosons in the theory is fit at $\sim 125~{\rm GeV}$. The signal strength 
measurements in different Higgs decay channels put further constraints on the Higgs sector, particularly on the 
composition of the SM-like Higgs boson. The 125 GeV Higgs 
with a sizable singlet component can result in a smaller production cross-section and/or branching ratio in its
SM decay modes. This would  produce weaker signal strength values in those channels, which are ruled 
out by the recent Higgs data.   Also the total decay width of this Higgs boson cannot be arbitrarily large. 
CMS puts an upper limit on the ratio of the total decay width of this Higgs boson, 
$\frac{\Gamma_{h_2}}{\Gamma_{h_{SM}}}< 4.15$\footnote{In  the $U(1)^\prime$ model, the lightest Higgs boson $h_1$ is 
lighter than the SM-like Higgs boson $h_2$.} \cite{cms-hdec}, which we include as another constraint. Any non-standard 
Higgs sector is highly constrained due to lack of any evidence of another Higgs boson other than the one which 
is mostly SM-like. However, all Higgs masses and their couplings are extremely important for this  
study,  as they appear as propagators in the pair annihilation process. Hence we consider all the Higgs 
measurement data including LEP, Tevatron and LHC to test our spectrum using {\tt HiggsBounds} \cite{Bechtle:2008jh}.

A light neutralino opens up a new (invisible) decay mode for the SM-like Higgs boson. But from the most recent measurements 
of various Higgs decay modes, any non-standard Higgs decay branching ratio is suppressed. For the search for a light neutralino, 
this is a very serious constraint, because if kinematically allowed, i.e, if $m_{\widetilde\chi^0_1}\le m_h /2$, the Higgs boson 
may now decay into a pair of LSP neutralinos, showing up as missing energy at the colliders. ATLAS and CMS have been separately 
searching for this non-standard Higgs decay mode in different Higgs boson production channels. The combined CMS limit on the invisible 
Higgs branching ratio (BR) derived from vector-boson fusion (VBF) and associated $Zh$ production modes is 
$58\%$ \cite{Chatrchyan:2014tja}, while the same limit derived from associated $Zh$ production mode by ATLAS is $75\%$ \cite{atlas-coup}. 
However, ATLAS puts a more stringent limit ($68\%$) from coupling measurements of the Higgs boson \cite{Aad:2014iia}. 
Global parameter fits to the data put this limit much lower, 28\% at 95\%C.L. \cite{Giardino:2013bma}. We chose to impose  
the direct search limit obtained from CMS for the present study.  

In the Table \ref{tbl:constraints} below, we show all the constraints which enter our calculations. 
\setlength{\voffset}{-0.5in}
\begin{table}[h!]
\centering
\begin{tabular}{||l | l||}
\hline
LEP mass limits  & $m_{\tilde \chi_1^\pm} >103.5$ GeV \\
$m_{h_2}$ &   $ [124.5, 126.5] $ GeV\\
Invisible Higgs width & $BR(h_2 \to \tilde \chi_1^0 \tilde \chi_1^0)<58$ \%\\
Invisible Z width & $\Gamma (Z \to \tilde \chi_1^0 \tilde \chi_1^0)<3$ MeV\\
Higgs width & $\frac{\Gamma_{h_2}}{\Gamma_{h_{SM}}}< 4.15$\\
$BR(B_s \to \mu^+ \mu^-)$ &  $[2.1 \times 10^{-9}, 2.9 \times 10^{-9}]$\\
$BR(b \to s \gamma )$ &  $[3.35 \times 10^{-4}, 3.7 \times 10^{-4}]$\\
Relic Density &$ [0.1118,0.128] $ \\
Direct Detection & LUX bounds \\
$\Delta a_{\mu}$ & $(26.1\pm 8.0)\times 10^{-10}$\\
$\Delta a_e$  & $(109\pm 83)\times 10^{-14}$ \\
\hline
\end{tabular}
\caption{Experimental Constraints implemented in the analysis. Here $h_2$ is the SM-like Higgs boson in the $U(1)^\prime$ model.}
\label{tbl:constraints}
\end{table}

For the analysis, we proceed as follows. We implement the $U(1)^\prime$ model into {\tt CalcHEP} \cite{Belyaev:2012qa} 
using the model implementation available in {\tt SARAH} \cite{Staub:2013tta}. For the calculation of the 
dark matter relic density we use {\tt micrOMEGAs} \cite{Belanger:2013oya}. Masses of the supersymmetric 
particles were evaluated using {\tt SPheno} \cite{Porod:2003um}, and we used {\tt HiggsBounds}  \cite{Bechtle:2008jh}
to impose the present collider limits on the Higgs mass and signal strengths at the LHC.

Scanning over the whole $U(1)^\prime$ parameter space, it is evident that only the gaugino and the Higgs mass and mixing 
parameters are crucial for our present study. The other SUSY parameters do not play a crucial role in determining DM observables, 
namely, for relic density, direct detection cross section and thermal averaged annihilation cross section.
Hence we chose to vary the low energy values of four relevant parameters randomly, in the ranges provided in 
Table~\ref{scan-range}.
\begin{table}[h!]
\begin{center}
\begin{tabular}{||c|c||}\hline\hline 
Parameters & Scan Ranges    \\ 
\hline\hline
$v_S~({\rm GeV})$ & (100 , 1000)    \\
$\lambda$ & (0.1 , 0.9) \\
${\rm tan}\beta $ & (2 , 50) \\
$M_1^{\prime}~({\rm GeV})$ & (100 , 1000)  \\
$A_t~({\rm GeV})$ & (-2500 , -500)  \\
$A_S~({\rm GeV})$ & (100 , 1000)  \\
\hline\hline 
\end{tabular}
\end{center}
\caption{Scan ranges of the randomly varied model parameters for light DM search.}
\label{scan-range}
\end{table}

The other parameters are set at the following values throughout the analysis:
\begin{eqnarray}
M_1 = 1~{\rm TeV};~~~~M_2 = 2~{\rm TeV};~~~~M_3 = 2.5~{\rm TeV};~~~~m_{\tilde f}\simeq 2~{\rm TeV},
\label{param-soft}
\end{eqnarray}
where $m_{\tilde f}$ are the soft masses of the squarks and the sleptons for all three generations. These parameters 
were chosen to highlight the effects of the $U(1)^\prime$ symmetry.
\subsection{Results}
\label{subsec:results}
\subsubsection{Relic Density and direct detection cross-section}
\label{subsubsec:relic}
The lightest neutralino in this scenario emerges as  dominantly singlino-type with a small admixture of  
higgsino component. Since for the present study we are only interested in a DM candidate with mass below $50~{\rm GeV}$, 
 we do not take into account points which yield a heavier DM mass. 
For this mass range, the DM pair dominantly annihilates into a quark-antiquark  pair. The process is mediated 
by neutral CP-even and CP-odd Higgs bosons and by $Z$ and $Z^{\prime}$ gauge bosons. Since in our considerations 
the sfermions are substantially heavier, their co-annihilation with the DM is suppressed; also since the LSP is 
very light compared to other supersymmetric particles, co-annihilation with other neutralinos and with the charginos 
is negligible.   

As $\mu_{eff}$ is given by the product of $\lambda$ and $v_S$, both these parameters play a crucial 
role in determining the lighter chargino mass, which for our choice of parameters is mostly higgsino. Therefore, 
imposing the chargino mass constraint provided by the LEP prevents $\mu_{eff}$ to be arbitrarily small and thereby reduces the 
freedom of choosing the gaugino mass spectrum. A light Higgs mass below 100 GeV mark is tightly constrained 
from the LEP Higgs boson search. Starting from an electron-positron initial state  at LEP, a too light Higgs can 
result in a larger cross-section for $Zh$ final state \cite{Barate:2003sz}. We discard any such 
points in the parameter space using {\tt HiggsBounds}. 

\begin{figure}[h!]
\centering
\includegraphics[height=6cm,width=7cm]{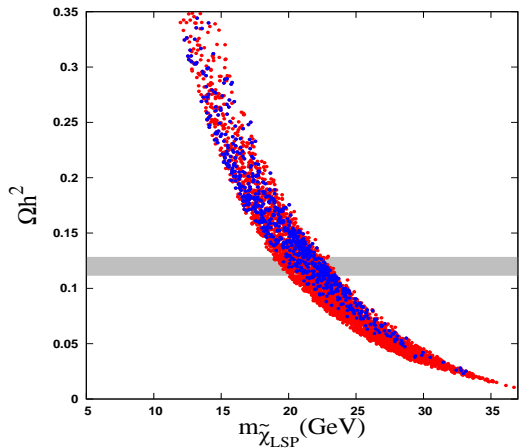}
\caption{(color on-line) Relic density distribution as a function of the neutralino DM mass. The horizontal band represents 
$3\sigma$ allowed range of relic density: $0.128\ge\Omega h^2\ge 0.1118$. The red 
points show the general features of the distribution and the blue points are obtained after 
imposing the {\tt HiggsBounds} constraint in addition to $BR(h_{2_{\rm inv}}) < 58\%$, where $h_2$ is the SM-like Higgs 
boson in the theory.} 
\label{fig:1}
\end{figure}

In Fig.~\ref{fig:1} we show the relic density distribution as a function of the DM mass. The shaded horizontal 
line represents $3\sigma$ allowed region around the best-fit relic density value provided by PLANCK. The red 
points show the general features of the distribution and the blue points in the plot satisfy {\tt HiggsBounds} 
constraint and also have a Higgs invisible decay branching ratio below $58\%$. As evident from the plot, 
the neutralino mass can be as light as $20~{\rm GeV}$ in this scenario. Below this threshold, annihilation 
of the DM is not sufficient to produce the correct relic density. In principle,  a lighter DM mass is possible 
if the lightest CP-even Higgs mass is light enough to enhance the annihilation cross-section, but that requires
increasing the singlet component in the lightest Higgs state which modifies the mixing in the Higgs sector and 
affects the signal strengths measured at the LHC. We return to a discussion of the consequences  of this scenario 
later in this section. 

\begin{figure}[h!]
\centering
\includegraphics[height=6cm,width=7cm]{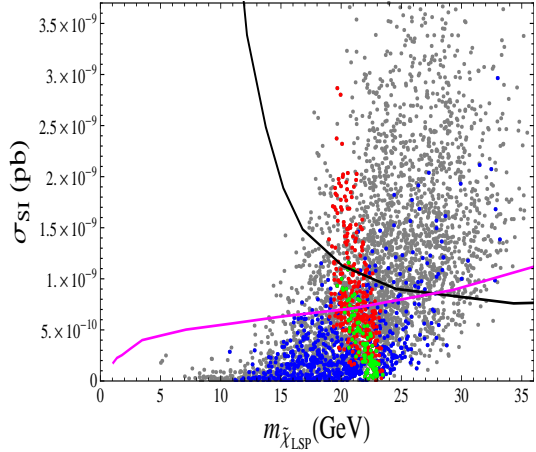}
\caption{(color on-line) Spin-independent cross-section as a function of the neutralino DM mass. The black line and magenta 
lines represent most recent exclusion lines provided by LUX and ATLAS experiments, respectively. The gray points show the 
general features, the red (and blue) points satisfy DM (and Higgs) constraints while the green ones satisfy both simultaneously.} 
\label{fig:2}
\end{figure}

From the direct detection experiments, experimental sensitivity to spin-independent (SI) scattering is much 
larger than to spin-dependent since spin-independent processes scatter coherently, and therefore are enhanced in 
scattering from large target nuclei. The most stringent bound on the spin-independent $\sigma_{SI}$ cross section 
in terms of the DM mass is provided by ATLAS \cite{atlas-coup}, more stringent than the corresponding limit provided 
by DM experiments at low mass region. Therefore, we take this constraint and the one from the LUX experiment 
\cite{Akerib:2013tjd} into account  in Fig.~\ref{fig:2}, where we plot the spin-independent DM cross section from 
direct searches as a function of the DM mass. 

The black line in Fig.~\ref{fig:2} represents the LUX bound whereas the 
magenta line shows the ATLAS limit. The gray points show the general features, the red points satisfy relic 
density constraint, the blue points satisfy invisible decay branching ratio and {\tt HiggsBounds} constraints. 
The green points satisfy both these and the other collider constraints simultaneously.
Contributions to $\sigma_{SI}$ coming from $t$-channel scalar exchange  diagrams 
are suppressed by a factor of $\frac{1}{m^4_{h_i}}$. Hence a light CP-even Higgs in the theory can 
enhance the scattering cross-section compared to the usual MSSM scenario. As the LSP neutralino has a 
large singlino component, it couples strongly to the light CP-even Higgs which is dominantly singlet-like. 
As a result, the scattering cross-section is large in this scenario, as can be seen from Fig.~\ref{fig:2}. The ATLAS 
limit reduces the parameter space further than the present LUX result. The remaining parameter space may be 
completely ruled out from future runs of LUX and XENON1T \cite{Aprile:2012zx}.   
\subsubsection{Annihilation cross-section}
\label{subsubsec:annihilation}
\begin{figure}[h!]
\centering
\includegraphics[height=6cm,width=7cm]{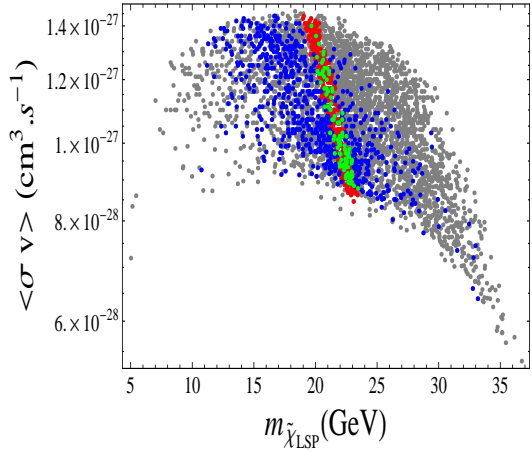}
\caption{(color on-line) Annihilation cross-section as a function of the neutralino DM mass. The color codes are same as before. Gray 
points show the general feature, blue points satisfy {\tt HiggsBounds} and invisible decay width constraints, red points 
satisfy the relic density constraint and the green ones satisfy both these and other collider constraints simultaneously.}
\label{fig:3}
\end{figure}

A light CP-even Higgs boson appearing as a mediator enhances the DM-pair annihilation cross-section into 
fermion pairs. As the coupling between the neutralino pairs and SM particles ($Z$ and Higgs bosons)
needs a suitable value to give an acceptable annihilation rate, the lightest neutralino is required to have non-zero 
SM-components. For a singlino-dominated neutralino, the annihilation proceeds mostly through $h_1$, while for a 
higgsino-dominated neutralino, the annihilation is dominated by the $Z$ boson. For the light neutralino in our model, 
the DM annihilation cross section obtained is at most $\sim 10^{-27}~{\rm cm^3\cdot s^{-1}}$. This annihilation 
cross section, as shown in Fig.~\ref{fig:3},  besides satisfying direct detection constraints,
is sufficient to produce correct relic density.  The color codes used in this figure are same as in Fig.~\ref{fig:2}.  

In the present scenario, the annihilation cross-section cannot be enhanced to ${\cal O}(10^{-26}~{\rm cm^3 \cdot s^{-1}})$ 
limit for a 30-40 GeV DM mass, as required to explain the galactic center gamma-ray excess observed by Fermi-LAT experiment 
\cite{Daylan:2014rsa}. One must fit the lightest CP-even Higgs or the pseudoscalar mass in the 60-80 GeV region 
in order to produce a resonating effect in the pair annihilation process to enhance the cross-section. In that case, the 
lightest CP-even Higgs must be dominantly singlet-like as a non-negligible doublet Higgs component may increase its production 
rate and violate the LEP limit \cite{Barate:2003sz}. On the other hand, with light CP-even or pseudoscalars in the model, 
one needs to be careful with its coupling to the SM-like boson because, if kinematically allowed, the $\sim 125$ GeV Higgs 
may decay into a pair of these scalars. Non-standard decays like this are highly constrained from Higgs signal strength 
measurements. Since our motivation is to look for the lightest neutralino DM  which lies very close to 20 GeV, we did not 
try to fit this enhanced annihilation cross-section for the present analysis. For our scenario, the pseudoscalar mass is 
kept above 1 TeV throughout and the lightest CP-even Higgs boson mass lies above $80~{\rm GeV}$ mark. 
\subsubsection{Composition of the LSP}
\label{subsubsec:composition}
\begin{figure}[h!]
\centering
\includegraphics[height=6cm,width=7cm]{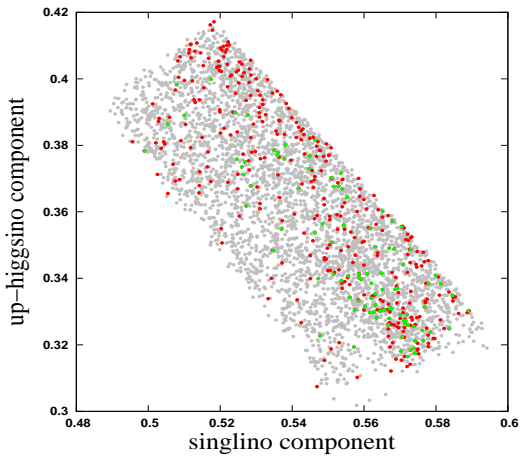}
\caption{(color on-line) The dominant components of the LSP neutralino, the singlino component on the $x$-axis, and the   
 up-type higgsino on the $y$-axis. The LSP must be dominantly singlino, as explained in the text. The gray points show 
the general features, the red points satisfy DM constraints, while the green ones satisfy both the collider and DM constraints.} 
\label{fig:4}
\end{figure}

The neutralino LSP, though mainly singlino-like,  has also a non-negligible higgsino component. This  
cannot be too large because the neutralino pair couples with the SM-like Higgs mainly through this component. 
Although the ${\cal O}(125~{\rm GeV})$ Higgs in this model also has a singlet component, it has to be very small so as to 
make it SM-like. There is a very delicate balance in the singlet admixture in both the lightest Higgs boson and the LSP 
neutralino if one must satisfy all constraints. Increasing the singlet VEV  $v_S$, one can increase the singlino component 
in the LSP neutralino. But this in turn increases the mixing in Higgs sector. As a result, the lightest CP-even Higgs is no longer 
dominantly singlet-like and the value of its coupling to the neutralino pair is reduced. Hence one has to have 
a neutralino DM with heavier (than $20~{\rm GeV}$) mass in order to produce correct relic density. Also, in this situation, the 
${\cal O}(125~{\rm GeV})$ Higgs boson has a significant component  from the singlet state which increases its invisible 
decay branching ratio as well as reducing its coupling to the SM particles. Hence one cannot have a too large singlino component 
in the LSP state. In Fig.~\ref{fig:4} we plot the two dominant contributions to the LSP state, coming from the singlino 
and the up-type higgsino. Red points satisfy DM constraints while the green ones satisfy all the constraints. This LSP is 
different from the neutralino LSP in both MSSM, where it is mostly bino \cite{MSSM}, or in NMSSM, where it is either singlino, 
or a singlino-bino mixture \cite{NMSSM}. 
\subsubsection{Light Higgs bosons and light chargino}
\label{subsubsec:chargino}
\begin{figure}[h!]
\centering
\includegraphics[height=6cm,width=7cm]{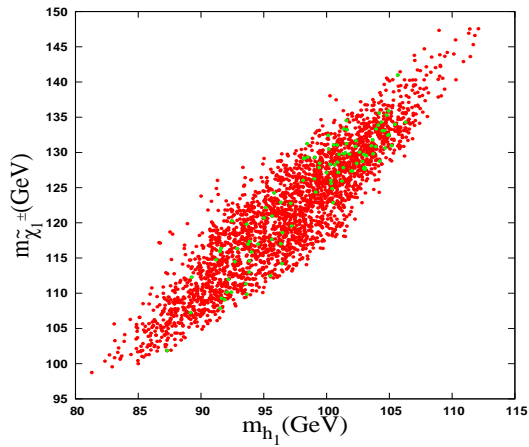}
\caption{(color on-line) The lighter chargino mass plotted against the lightest Higgs mass in the model. 
Red points satisfy Higgs invisible branching ratio decay constraints while the green ones satisfy 
both collider and DM constraints.} 
\label{fig:5}
\end{figure}

As Fig.~\ref{fig:4} suggests, to satisfy both the DM and Higgs boson constraints, a singlino-higgsino mixed LSP 
neutralino state is required in this model. This imposes a constraint on the $\lambda$ parameter which, therefore, 
cannot be arbitrarily small. The lighter chargino in this model is higgsino-like and its mass increases with increasing 
$\lambda$. Note  that this is different from the usual MSSM scenario, where the lightest chargino is the wino. The lightest 
Higgs boson is non-SM-like and its mass is tightly constrained from the LEP data. But if it is completely singlet-like, 
or singlet with a small admixture of doublet Higgs, then the bound can be relaxed. In our scenario, the favored 
mass range for the lightest Higgs mass is $m_{h_1} \sim 85-105$ GeV. In Fig.~\ref{fig:5} we show the distribution 
of the lighter chargino mass ($m_{\widetilde\chi^{\pm}_1}$) and the lightest Higgs boson mass ($m_{h_1}$) 
in one slice of the parameter space that is favored from both DM and collider constraints (the green points). 
As before the red points satisfy Higgs invisible branching ratio decay constraints only. 
\subsubsection{Higgs signal strengths}
\label{subsubsec:LHC}
The Higgs sector is constrained from the ${\cal O} (125~{\rm GeV})$ 
Higgs boson signal strengths  measurements in different decay modes.
These Higgs measurements data provide the strongest limits on the neutralino DM mass in the model. 
The signal strength in different channel ($\mu_i$) is defined as 
\begin{equation}
\mu_i = R_{i}^{\rm prod}\times \frac{R_{i}^{\rm decay}}{R^{\rm width}}\, .
\end{equation}
Here $R_i$'s are $U(1)^\prime$ model predictions for the ratios of the Higgs production cross sections and 
partial decay rates for various channels, and $R^{\rm width}$ is the ratio of the total width in our model, defined with respect to the 
corresponding SM expectations.
\begin{eqnarray}
R_i^{\rm prod} = \frac{\left(\sigma_i^{\rm prod}\right)_{\rm U(1)'}}{\left(\sigma_i^{\rm prod}\right)_{\rm SM}}\, ,~
R_i^{\rm decay} = \frac{\left(\Gamma_i^{\rm decay}\right)_{\rm U(1)'}}{\left(\Gamma_i^{\rm decay}\right)
_{\rm SM}}\, ,~
R^{\rm width} = \frac{\left(\Gamma^{\rm width}\right)_{\rm U(1)'}}{\left(\Gamma^{\rm width}\right)_{\rm SM}}\, , 
\end{eqnarray}
We require that the SM-like Higgs boson in the $U(1)^\prime$ model must yield signal strength values lying within 
the $1\sigma$ error bars of the best-fit values obtained by CMS and ATLAS, as listed in Table~\ref{higgs-dat}, with 
the measured values denoted as $\widehat \mu_i$, while we keep the notation $\mu_i$ for the predicted signal strengths.
\begin{table}[htb]
\centering
\begin{tabular}{||c|c|c||}
\hline\hline
Channel & $\widehat{\mu}$ & Experiment \\
\hline \hline
  $h \to \gamma \gamma$ & $1.29^{+0.30}_{-0.30}$ & ATLAS~\cite{Aad:2014aba} \\
  & $1.14 ^{+0.26}_{- 0.23}$ & CMS~\cite{Khachatryan:2014ira} \\ \hline
  $h \rightarrow Z Z^{\star} \to 4l$ & $1.44^{+0.40}_{-0.35}$ & ATLAS~\cite{atlasmu} \\
 & $0.93_{-0.27}^{+0.27}$ & CMS~\cite{cmsz} \\ \hline
  $h \rightarrow W W^{\star} \to 2l 2\nu$ & $1.00_{-0.29}^{+0.32}$ & ATLAS~\cite{atlasmu} \\
& $0.72_{-0.18}^{+0.20}$ & CMS~\cite{cmsw} \\ \hline
  $h \rightarrow b \bar{b}$ & $0.20_{-0.60}^{+0.70}$ & ATLAS (VH)~\cite{atlasmu} \\
& $1.00_{-0.50}^{+0.50}$ & CMS (VH)~\cite{cmsfer}\\ \hline
  $h \rightarrow \tau \bar{\tau}$ & $1.4_{-0.4}^{+0.5}$ & ATLAS~\cite{atlasmu} \\
& $0.78_{-0.27}^{+0.27}$ & CMS~\cite{cmsfer} \\ \hline
\hline
\end{tabular}
\caption{Combined results at 7 and 8 TeV for the measured signal strength ($\widehat{\mu_i}$) values in various channels 
and their $1\sigma$ uncertainties as reported by the ATLAS and CMS collaborations.}
\label{higgs-dat}
\end{table} 

\begin{figure}[htbp]
\begin{center}$
    \begin{array}{cc}
\hspace*{-0.7cm}
\includegraphics[height=6cm,width=7cm]{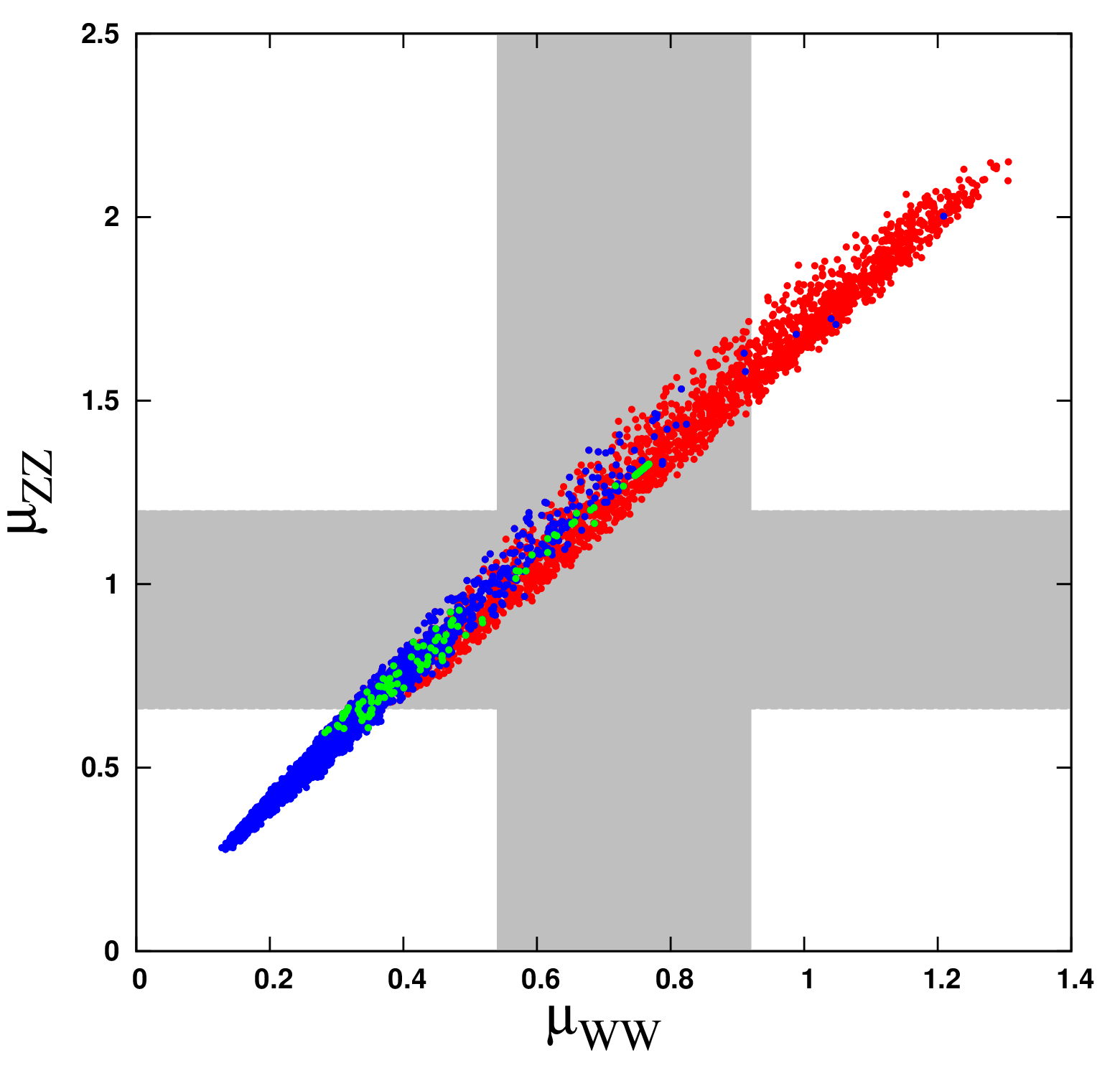} 
& \hspace*{1.5cm} \includegraphics[height=6cm,width=7cm]{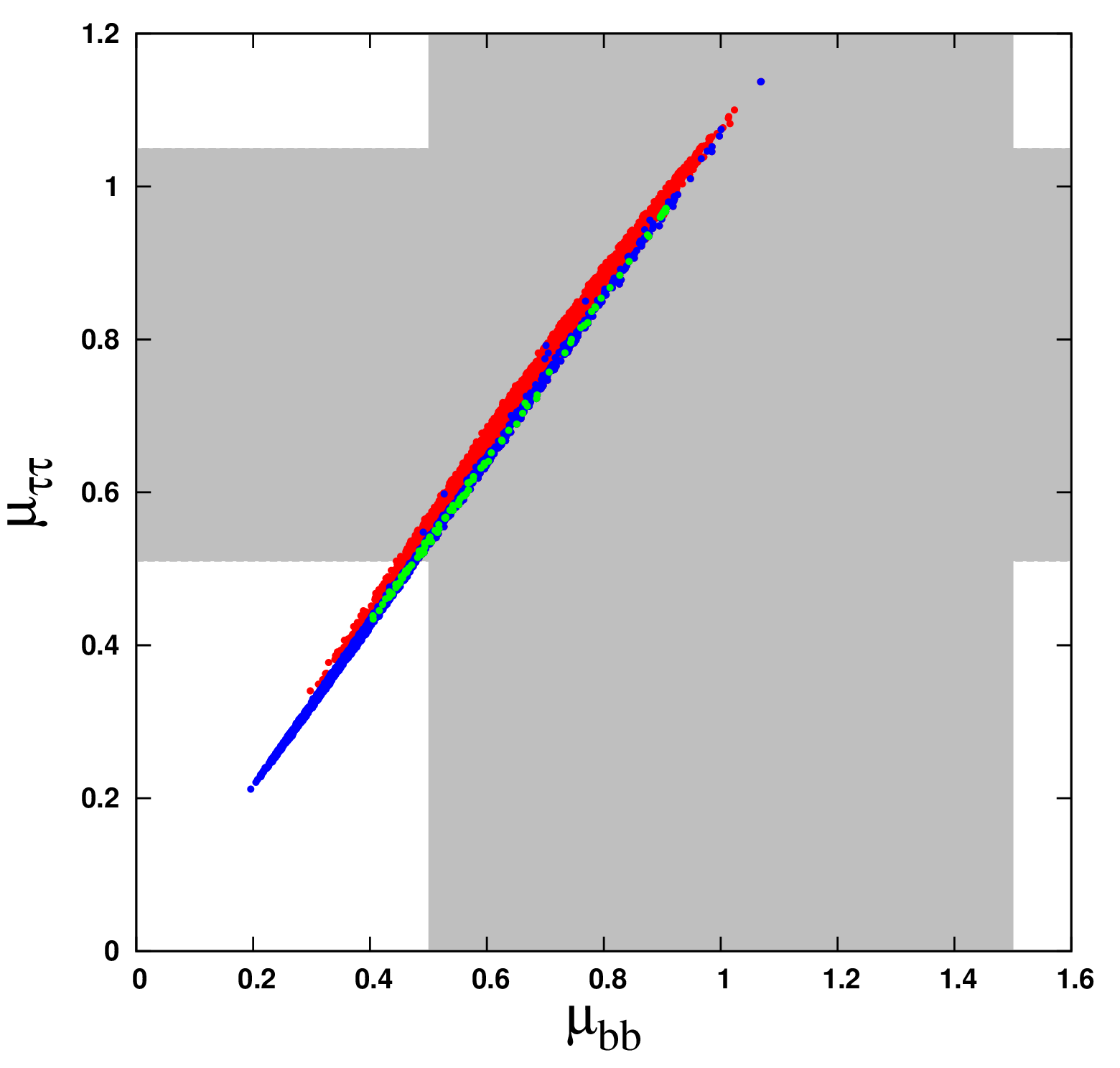}
 \end{array}$
\end{center}
\caption{(color on-line) Left panel: the Higgs boson signal strengths in gauge boson decay channels ($\mu_{WW}$ and 
$\mu_{ZZ}$) for the corresponding parameter space shown in the plot. Right panel: the Higgs boson signal strengths 
in fermion anti-fermion decay channels ($\mu_{bb}$ and $\mu_{\tau\tau }$) 
for the corresponding parameter space shown in the plot. The red points correspond to $BR(h_{2_{\rm inv}}) < 58\%$. The 
points which satisfy  {\tt HiggsBounds}  are shown in blue. The green points are the ones which satisfy all the constraints 
simultaneously. The shaded regions are the $1\sigma$ allowed ranges for the respective signal strength values as quoted 
by the CMS collaboration.}
\label{fig:6}
\end{figure}

To highlight our parameter space and its implications, we show some correlation plots between the signal strengths obtained 
for decays into different final states. Fig.~\ref{fig:6} shows on the left-hand side, the correlation between signal strengths 
in gauge boson decay channels $\mu_{WW}$ and $\mu_{ZZ}$ while the right-hand side of Fig.~\ref{fig:6} illustrates  
a similar correlation between fermion anti-fermion decay channels $\mu_{bb}$ and $\mu_{\tau\tau}$. 
For the calculation of $\mu_{bb}$, we consider gauge boson associated Higgs production channel $Vh$, while for all 
the other signal strength calculations  we use Higgs boson production through gluon fusion. The red 
points in the plots satisfy the Higgs invisible decay branching ratio constraint, whereas the blue points satisfy 
{\tt HiggsBounds} constraints. The green points satisfy both these constraints,  in addition to the DM and other 
collider bounds. The shaded regions are the $1\sigma$ allowed ranges for the respective signal strength values as quoted 
by the CMS.

\begin{figure}[h!]
\centering
\includegraphics[height=6cm,width=7cm]{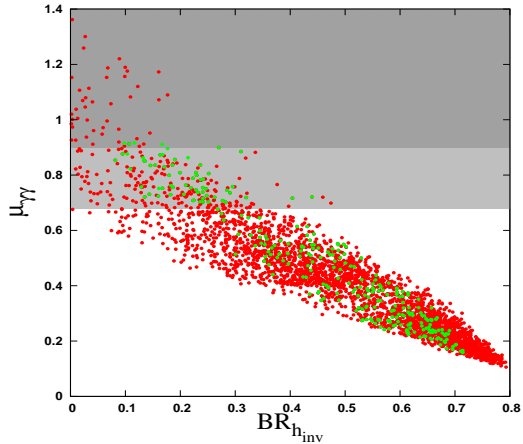}
\caption{(color on-line) Correlation between $BR(h_{2_{\rm inv}})$ and $\mu_{\gamma\gamma}$. The red 
points satisfy {\tt HiggsBounds} constraints and the green points simultaneously 
satisfy all the collider and DM constraints. The dark shaded region is the $1\sigma$ allowed range and 
the light shaded region is the $2\sigma$ allowed range for the 
$\gamma \gamma$ signal strength values as quoted by the CMS collaboration.}
\label{fig:7}
\end{figure}

In Fig.~\ref{fig:7} we plot $\mu_{\gamma\gamma}$ as a function of Higgs invisible branching ratio. 
A smaller $BR(h_{2_{\rm inv}})$ enhances $\mu_{\gamma\gamma}$ as expected. But in our scenario, the  
$BR(h_{2_{\rm inv}})$ cannot be very small as the LSP neutralino has a higgsino component which 
cannot be neglected. The largest $\mu_{\gamma\gamma}$ that we obtain after imposing all the constraints 
lies close to 0.92 which is just above $1\sigma$ lower limit of the value obtained by CMS. In Fig.~\ref{fig:7}, 
the dark shaded region shows $1\sigma$ reach and the light shaded region shows $2\sigma$ reach of the 
measured CMS value of $\mu_{\gamma\gamma}$. It shows that although our scenario is in slight tension 
with the best-fit CMS value, it still accommodates the $1\sigma$ experimental range, while if we take the $2\sigma$ allowed range a sizeable parameter space is   
still allowed. 

It is evident from Figs.~\ref{fig:6} and \ref{fig:7} that in this scenario, apart from $\mu_{ZZ}$, 
all the signal strengths values are closer to those measured by the CMS rather than those measured by the ATLAS.  
We observe that the two-photon signal strength can be enhanced close to the CMS best-fit value 1.14, but in that case, the 
DM constraints are hard to satisfy. Increasing the two-photon decay also means increasing the other SM decay BRs of the 
SM-like Higgs boson. Enhancement of these other decay modes may produce excess in some quark or leptonic final states 
that has not yet been observed and thus ruled out by {\tt HiggsBounds}. Also, note that increasing the signal strength in 
SM decay modes means decreasing  the invisible decay branching ratio, which means reduced coupling between the Higgs boson and 
the DM neutralino pair. As a result the DM pair annihilation may not be sufficient to produce correct relic density.   
\subsection{Three Benchmark points}
\label{subsec:BPs}
In this section, we present three benchmark points consistent with the previous discussions.  For each, we also quote the 
values for the relevant constraints and the SM-like Higgs signal strengths in various channels. 
The input values of the randomly varied parameters for these benchmark points are shown in Table~\ref{bp-input}. 
Although these benchmarks are obtained as random variations, the values for the relevant parameters are very close, 
indicating the fact that the $U(1)^\prime$ model parameter range is significantly restricted by the  cosmological and 
collider constraints, and thus quite predictive. Note in particular that the singlet VEV is $v_S \sim 300$ GeV,  
$\tan \beta \sim {\cal O}(5-6)$ and the $U(1)^\prime$ gaugino mass is light,  $M_1^{\prime} \sim 300$ GeV. As we have 
chosen other gaugino masses in the TeV range, this makes the $\tilde Z^\prime$ the lightest gaugino. Also, the sign of 
the coupling $\lambda$ indicate that $sign(\mu_{eff})>0$ is preferred. 
\begin{table}[h!]
\begin{center}
\begin{tabular}{||c|c|c|c|c||}\hline\hline 
Parameter & BP1 & BP2 & BP3 \\ 
\hline\hline
$v_S~({\rm GeV})$ & 291.09  & 273.42  & 297.70 \\
$\lambda$ & 0.612  & 0.611  & 0.631 \\
${\rm tan}\beta $ & 5.57  & 5.51  & 5.30 \\
$M_1^{\prime}~({\rm GeV})$ & 315.0  & 289.54  & 301.35 \\
$A_t~({\rm GeV})$ & -724.97  & -696.62  & -633.78 \\
$A_S~({\rm GeV})$ & 581.72  & 542.65  & 587.50 \\
\hline\hline 
\end{tabular}
\end{center}
\caption{Input values of the randomly varied parameters for the four benchmark points. Other parameters are taken 
as in Eq.~\ref{param-soft}.}
\label{bp-input}
\end{table}

The corresponding values of the constraints for these benchmark points are listed in Table~\ref{bp-out}.
\begin{table}[h!]
\begin{tabular}{||p{4cm}|c|c|c|c||} \hline\hline
Observables & BP1 & BP2 & BP3 \\ \hline\hline
$m_{h_1}$ (GeV) & 102.87   & 97.21  & 106.32 \\ 
$m_{h_2}$ (GeV) & 124.79  & 124.72  &   124.97\\ 
$m_{\widetilde\chi_1^0}$ (GeV) & 21.78  & 21.05  & 22.91 \\
$m_{\widetilde\chi_1^{\pm}}$ (GeV) & 127.45  & 119.57  & 134.22 \\ \hline
$\Omega_{\rm DM}h^2$ & 0.128  & 0.126  &  0.119 \\ 
$\sigma_{\rm SI}$ (pb) & $2.89\times 10^{-10}$ & $4.86\times 10^{-10}$ & $1.65\times 10^{-10}$ \\ 
$<\sigma v>$ ($cm^3.s^{-1}$) & $9.67\times 10^{-28}$ & $1.11\times 10^{-27}$ & $8.83\times 10^{-28}$ \\ 
\hline
$\delta a_\mu$ & $2.10\times 10^{-11}$ & $1.98\times 10^{-11}$ & $2.02\times 10^{-11}$ \\
$\delta a_e$ & $4.80\times 10^{-16}$ &  $4.52\times 10^{-16}$ & $4.62\times 10^{-16}$ \\ \hline
BR$(B\to X_s\gamma)$ & $3.37\times 10^{-4}$ & $3.98\times 10^{-4}$ & $3.37\times 10^{-4}$ \\
BR$(B_s\to \mu^+\mu^-)$ & $2.71\times 10^{-9}$ & $2.63\times 10^{-9}$ & $2.74\times 10^{-9}$ \\ \hline
$BR(h_2\rightarrow \widetilde \chi_1^0 \widetilde \chi_1^0)$ & 0.13  & 0.20  & 0.09 \\
$\mu_{\gamma\gamma}$ & 0.92  & 0.88  & 0.91 \\
$\mu_{b\bar b}$ & 0.89  & 0.82  & 0.95 \\
$\mu_{WW}$ & 0.77  & 0.72  & 0.78 \\
$\mu_{ZZ}$ & 1.29  & 1.22  & 1.32 \\
$\mu_{\tau\tau}$ & 0.97  & 0.88  & 1.02 \\  \hline\hline
\end{tabular}
\caption{Values for the Higgs mass, DM mass, relic density, spin-independent cross section and   
low-energy flavor sector observables, along with the SM-like Higgs boson signal strengths in different decay modes, 
for the three chosen benchmarks in Table~\ref{bp-input}.}
\label{bp-out}
\end{table}

The lightest Higgs boson is mostly singlino and thus would have escaped detection at LEP, Tevatron or LHC. 
The second lightest Higgs boson is SM-like, and its signal strengths are compatible with the LHC data. Note in particular 
the predicted suppression of the $h \rightarrow \gamma \gamma$ signal, which is $1\sigma$ away from the best-fit value 
of the CMS measurement, and in slight conflict with the one at ATLAS. The invisible decay width is small and consistent 
with both experimental constraints and with the global fits to the LHC data \cite{Giardino:2013bma}. 
The LSP mass is $\sim 20$ GeV, while the lightest chargino (higgsino for our parameter choices) has mass $\sim 120-135$ GeV.  

\newpage
\section{Conclusion}
\label{sec:conclusion}
In this work, we investigated the possibility that a very light neutralino LSP in $U(1)^\prime$ models, 
consistent with constraints coming from cosmological, astrophysical and particle accelerator data, can be a viable 
candidate for cold dark matter. In particular, this neutralino must be compatible with relic density measurements, 
and with direct and indirect detection properties compatible with the recent data. Due to differences in the neutralino 
and Higgs sectors, the properties of the LSP in $U(1)^\prime$ models can be significantly distinct from those in 
the MSSM, both in its nature and in the processes relevant for the LSP relic density and for its detection. Unlike MSSM 
and NMSSM, DM annihilation can occur through $Z^\prime$ mediated channels, and the LSP neutralino could have a 
$\tilde Z^\prime$ component. However, we find that, in order for the LSP neutralino to be very light, it must be singlino-like. 
The relevant model parameters are the neutralino and Higgs mass and mixing parameters. We vary the VEV of the singlet field, 
the mass of the $\tilde Z^\prime$, the ratio of the doublet VEVs, the trilinear soft parameters for the stop and the singlet, 
and the coupling constant of the triple Higgs interaction term (responsible for generating $\mu_{eff}$). The other gaugino masses, corresponding to 
$SU(3)_c,~SU(2)_L$,  $U(1)_Y$, as well as masses for all sfermions, are in the TeV range.  

The best fit is obtained by an LSP neutralino which is dominantly singlino-like, with a small admixture of up-type higgsino. 
This is unlike in the MSSM, where a light LSP is the $U(1)_Y$ bino, or in the NMSSM, where it is a mixture of $U(1)_Y$ 
bino and singlino. The additional gauge and Higgs bosons can contribute to the LSP annihilation and detection processes.  
The singlino mixed with an up-higgsino  LSP state can satisfy all 
present cosmological constraints. In addition,  imposing stringent constraints on the DM mass and the annihilation 
cross-section from the Higgs data at the LHC severely restricts the non-SM-like lightest Higgs boson, required to 
agree with LEP bounds, as well as non-SM components of the SM-like Higgs boson required to satisfy signal strengths measurements 
at the LHC. We would like to note that, should signal strengths not be included, our parameter space would be much 
less restricted and  lighter LSP neutralino masses can be obtained. After imposing {\it all} the collider and DM 
constraints, one can obtain a neutralino DM candidate 
as light as 20 GeV. We take into account both the ATLAS and CMS measured signal strength values for different 
Higgs boson decay modes and calculate them in the parameter region favored by our study. We observe that 
although we can fit the CMS  experimental values within their $1 \sigma$ errors, there is  tension with the ATLAS measurements. 
In addition to scanning over the relevant parameter space, we also provide as examples, three concrete benchmark 
points and list the corresponding values of the constraints that were taken into account.    

The allowed parameter values selected by scanning over the range of relevant $U(1)^\prime$ model parameters are very 
restrictive and thus very predictive. In particular, the lightest CP-even neutral Higgs boson is mostly singlet, 
while the SM-like Higgs boson contains some admixture of singlet. The lightest chargino is  higgsino-like, and the 
lightest gaugino is the $U(1)^\prime$ bino. In most cases there are four neutralinos within the energy range observable 
at the LHC, with production and cascades decays similar to the MSSM. However, all of these would undergo a further 
decay to a singlino LSP, accompanied e.g., by an on-shell $Z$ or Higgs bosons. 
A signature for these neutralinos is a  trilepton signal  different from that in the MSSM, as the masses and couplings of the 
neutralinos are modified. Enhanced rates for the decay of neutralino pairs into three or more leptons are possible, yielding 
events with five (or more) lepton signals. As the squarks and gluinos which can give similar signals are heavy in this model, 
the multi-lepton  signature will mostly indicate the singlino-dominated LSP. Future experimental constraints are expected from 
LHC at 14 TeV monojet with luminosities 100 fb$^{-1}$ and 300 fb$^{-1}$, XENON1T, and AMS-02 one year antiproton data. 
As the light neutralino solution requires a restricted parameter space, and a delicate interplay among compositions of the LSP, 
the lightest CP-even, and the SM-like Higgs bosons,  these experiments will be instrumental in further confirm, restrict or 
rule it out.  
\section{Acknowledgement}
The work of M.F. is supported in part by NSERC under grant number SAP105354 and would like to acknowledge the 
hospitality of IACS, Kolkata, where this work was started. 
S.M. wishes to thank Sourov Roy for some useful discussions and the Department of Science and Technology, Government of India, for 
a Senior Research Fellowship.  




\begin{thebibliography}{99}
\bibitem{Chatrchyan:2012ufa}
  S.~Chatrchyan {\it et al.}  [CMS Collaboration],
  Phys.\ Lett.\ B {\bf 716} (2012) 30
  [arXiv:1207.7235 [hep-ex]].
  
\bibitem{Aad:2012tfa}
  G.~Aad {\it et al.}  [ATLAS Collaboration],
  Phys.\ Lett.\ B {\bf 716} (2012) 1
  [arXiv:1207.7214 [hep-ex]].
  
\bibitem{Komatsu:2010fb} 
  E.~Komatsu {\it et al.}  [WMAP Collaboration],
  Astrophys.\ J.\ Suppl.\  {\bf 192}, 18 (2011)
  [arXiv:1001.4538 [astro-ph.CO]].
  
\bibitem{Ade:2013zuv} 
  P.~A.~R.~Ade {\it et al.}  [Planck Collaboration],
  arXiv:1303.5076 [astro-ph.CO].

\bibitem{Bernabei:2008yi} 
  R.~Bernabei {\it et al.}  [DAMA Collaboration],
  Eur.\ Phys.\ J.\ C {\bf 56}, 333 (2008)
  [arXiv:0804.2741 [astro-ph]].
  
\bibitem{Aalseth:2010vx} 
  C.~E.~Aalseth {\it et al.}  [CoGeNT Collaboration],
  Phys.\ Rev.\ Lett.\  {\bf 106}, 131301 (2011)
  [arXiv:1002.4703 [astro-ph.CO]].
  C.~E.~Aalseth, P.~S.~Barbeau, J.~Colaresi, J.~I.~Collar, J.~Diaz Leon, J.~E.~Fast, N.~Fields and T.~W.~Hossbach {\it et al.},
  Phys.\ Rev.\ Lett.\  {\bf 107}, 141301 (2011)
  [arXiv:1106.0650 [astro-ph.CO]].
  
\bibitem{Agnese:2013rvf} 
  R.~Agnese {\it et al.}  [CDMS Collaboration],
  Phys.\ Rev.\ Lett.\  {\bf 111}, 251301 (2013)
  [arXiv:1304.4279 [hep-ex]].
  
\bibitem{Angloher:2011uu} 
  G.~Angloher, M.~Bauer, I.~Bavykina, A.~Bento, C.~Bucci, C.~Ciemniak, G.~Deuter and F.~von Feilitzsch {\it et al.},
  Eur.\ Phys.\ J.\ C {\bf 72}, 1971 (2012)
  [arXiv:1109.0702 [astro-ph.CO]].
  
\bibitem{Aprile:2012nq} 
  E.~Aprile {\it et al.}  [XENON100 Collaboration],
  Phys.\ Rev.\ Lett.\  {\bf 109}, 181301 (2012)
  [arXiv:1207.5988 [astro-ph.CO]].
  
\bibitem{Adriani:2008zr} 
  O.~Adriani {\it et al.}  [PAMELA Collaboration],
  Nature {\bf 458}, 607 (2009)
  [arXiv:0810.4995 [astro-ph]].
  O.~Adriani, G.~C.~Barbarino, G.~A.~Bazilevskaya, R.~Bellotti, M.~Boezio, E.~A.~Bogomolov, L.~Bonechi and M.~Bongi {\it et al.},
  Phys.\ Rev.\ Lett.\  {\bf 102}, 051101 (2009)
  [arXiv:0810.4994 [astro-ph]].
  
\bibitem{Akerib:2013tjd} 
  D.~S.~Akerib {\it et al.}  [LUX Collaboration],
  Phys.\ Rev.\ Lett.\  {\bf 112}, 091303 (2014)
  [arXiv:1310.8214 [astro-ph.CO]].
  
\bibitem{FermiLAT:2011ab} 
  M.~Ackermann {\it et al.}  [Fermi LAT Collaboration],
  Phys.\ Rev.\ Lett.\  {\bf 108}, 011103 (2012)
  [arXiv:1109.0521 [astro-ph.HE]].
  
\bibitem{Ackermann:2011wa}
  M.~Ackermann {\it et al.}  [Fermi-LAT Collaboration],
  Phys.\ Rev.\ Lett.\  {\bf 107} (2011) 241302
  [arXiv:1108.3546 [astro-ph.HE]].
  
\bibitem{Boyarsky:2014jta} 
  A.~Boyarsky, O.~Ruchayskiy, D.~Iakubovskyi and J.~Franse,
  arXiv:1402.4119 [astro-ph.CO].
  
\bibitem{Bulbul:2014sua}
  E.~Bulbul, M.~Markevitch, A.~Foster, R.~K.~Smith, M.~Loewenstein and S.~W.~Randall,
  Astrophys.\ J.\  {\bf 789} (2014) 13
  [arXiv:1402.2301 [astro-ph.CO]].
  
\bibitem{Hooper:2002nq} 
  D.~Hooper and T.~Plehn,
  Phys.\ Lett.\ B {\bf 562}, 18 (2003)
  [hep-ph/0212226].

\bibitem{Bottino:2002ry} 
  A.~Bottino, N.~Fornengo and S.~Scopel,
  Phys.\ Rev.\ D {\bf 67}, 063519 (2003)
  [hep-ph/0212379].
  
\bibitem{Bottino:2003iu} 
  A.~Bottino, F.~Donato, N.~Fornengo and S.~Scopel,
  Phys.\ Rev.\ D {\bf 68}, 043506 (2003)
  [hep-ph/0304080].
  
\bibitem{Choudhury:2012tc} 
  A.~Choudhury and A.~Datta,
  JHEP {\bf 1206}, 006 (2012)
  [arXiv:1203.4106 [hep-ph]]; A.~Choudhury and A.~Datta,
  JHEP {\bf 09}, 119 (2013)
  [arXiv:1305.0928 [hep-ph]].
  
  \bibitem{MSSM}
  L.~Calibbi, J.~M.~Lindert, T.~Ota and Y.~Takanishi,
  arXiv:1405.3884 [hep-ph].
  K.~Hagiwara, S.~Mukhopadhyay and J.~Nakamura,
  Phys.\ Rev.\ D {\bf 89}, 015023 (2014)
  [arXiv:1308.6738 [hep-ph]].
  P.~Huang and C.~E.~M.~Wagner,
  arXiv:1404.0392 [hep-ph].
  G.~Belanger, S.~Biswas, C.~Boehm and B.~Mukhopadhyaya,
  JHEP {\bf 1212}, 076 (2012)
  [arXiv:1206.5404 [hep-ph]].
  G.~Bélanger, G.~Drieu La Rochelle, B.~Dumont, R.~M.~Godbole, S.~Kraml and S.~Kulkarni,
  Phys.\ Lett.\ B {\bf 726}, 773 (2013)
  [arXiv:1308.3735 [hep-ph]].
  C.~Boehm, P.~S.~B.~Dev, A.~Mazumdar and E.~Pukartas,
  JHEP {\bf 1306}, 113 (2013)
  [arXiv:1303.5386 [hep-ph]].
  T.~Han, Z.~Liu and A.~Natarajan,
  JHEP {\bf 1311}, 008 (2013)
  [arXiv:1303.3040 [hep-ph]].
  
  
  
  \bibitem{NMSSM}
  J.~Cao, C.~Han, L.~Wu, P.~Wu and J.~M.~Yang,
  JHEP {\bf 1405}, 056 (2014)
  [arXiv:1311.0678 [hep-ph]].
  J.~Kozaczuk and S.~Profumo,
  Phys.\ Rev.\ D {\bf 89}, 095012 (2014)
  [arXiv:1308.5705 [hep-ph]].
  K.~Ishikawa, T.~Kitahara and M.~Takimoto,
  arXiv:1405.7371 [hep-ph].
  T.~Han, Z.~Liu and S.~Su,
  arXiv:1406.1181 [hep-ph].
  
\bibitem{CMS:yva}
  [CMS Collaboration],
  CMS-PAS-HIG-13-005.
    
     
\bibitem{ATLAS:2013sla} 
  [ATLAS Collaboration],
  ATLAS-CONF-2013-034.

  

  
  
\bibitem{Cvetic:1995rj}
  M.~Cvetic and P.~Langacker,
  Phys.\ Rev.\ D {\bf 54} (1996) 3570
  [hep-ph/9511378].
  
\bibitem{Demir:2005ti}
  D.~A.~Demir, G.~L.~Kane and T.~T.~Wang,
  Phys.\ Rev.\  D {\bf 72} (2005) 015012
  [arXiv:hep-ph/0503290].


\bibitem{Cvetic:1997ky}
M.~Cvetic, D.~A.~Demir, J.~R.~Espinosa, L.~L.~Everett and
P.~Langacker,
Phys.\ Rev.\ D {\bf 56}, 2861 (1997) [Erratum-ibid.\ D {\bf 58},
119905 (1998)] [hep-ph/9703317].

\bibitem{muprob}
J.~E.~Kim and H.~P.~Nilles,
Phys.\ Lett.\ B {\bf 138}, 150 (1984);
D.~Suematsu and Y.~Yamagishi,
Int.\ J.\ Mod.\ Phys.\ A {\bf 10}, 4521 (1995)
[arXiv:hep-ph/9411239];
M.~Cvetic and P.~Langacker,
Mod.\ Phys.\ Lett.\ A {\bf 11}, 1247 (1996)
[arXiv:hep-ph/9602424];
V.~Jain and R.~Shrock,
arXiv:hep-ph/9507238;
Y.~Nir,
Phys.\ Lett.\ B {\bf 354}, 107 (1995) [arXiv:hep-ph/9504312].

\bibitem{Ellis:1986mq} 
  J.~R.~Ellis, K.~Enqvist, D.~V.~Nanopoulos, K.~A.~Olive, M.~Quiros and F.~Zwirner,
  Phys.\ Lett.\ B {\bf 176}, 403 (1986).


\bibitem{deCarlos:1997yv} 
  B.~de Carlos and J.~R.~Espinosa,
  Phys.\ Lett.\ B {\bf 407}, 12 (1997)
  [hep-ph/9705315].
  
\bibitem{Choi:2006fz} 
  S.~Y.~Choi, H.~E.~Haber, J.~Kalinowski and P.~M.~Zerwas,
  Nucl.\ Phys.\ B {\bf 778}, 85 (2007)
  [hep-ph/0612218].

  
\bibitem{Barger:2004bz} 
  V.~Barger, C.~Kao, P.~Langacker and H.~-S.~Lee,
  Phys.\ Lett.\ B {\bf 600}, 104 (2004)
  [hep-ph/0408120].
  
\bibitem{Barger:2007nv} 
  V.~Barger, P.~Langacker, I.~Lewis, M.~McCaskey, G.~Shaughnessy and B.~Yencho,
  Phys.\ Rev.\ D {\bf 75}, 115002 (2007)
  [hep-ph/0702036 [HEP-PH]]; 
  V.~Barger, P.~Langacker and G.~Shaughnessy,
  Phys.\ Lett.\ B {\bf 644}, 361 (2007)
  [hep-ph/0609068].
  
  
\bibitem{Demir:2009kc} 
  D.~A.~Demir, L.~L.~Everett, M.~Frank, L.~Selbuz and I.~Turan,
  Phys.\ Rev.\ D {\bf 81}, 035019 (2010)
  [arXiv:0906.3540 [hep-ph]].
  
\bibitem{Frank:2012ne} 
  M.~Frank, L.~Selbuz and I.~Turan,
  Eur.\ Phys.\ J.\ C {\bf 73}, 2656 (2013)
  [arXiv:1212.4428 [hep-ph]].

    
  
  
\bibitem{atlas:sparticle}
 ATLAS Collaboration, ATLAS-CONF-2013-035, ATLAS-CONF-2013-062, ATLAS-CONF-2013-092. 
\bibitem{cms:sparticle}
 CMS Collaboration, SUS-13-004, SUS-13-006, SUS-13-012.
 
\bibitem{Beringer:1900zz} 
  J.~Beringer {\it et al.}  [Particle Data Group Collaboration],
  Phys.\ Rev.\ D {\bf 86}, 010001 (2012).
 
\bibitem{Lees:2012ym} 
  J.~P.~Lees {\it et al.}  [BaBar Collaboration],
  Phys.\ Rev.\ Lett.\  {\bf 109}, 191801 (2012)
  [arXiv:1207.2690 [hep-ex]].
  
\bibitem{Aaij:2013aka} 
  R.~Aaij {\it et al.}  [LHCb Collaboration],
  Phys.\ Rev.\ Lett.\  {\bf 111}, 101805 (2013)
  [arXiv:1307.5024 [hep-ex]].

\bibitem{Chatrchyan:2013bka} 
  S.~Chatrchyan {\it et al.}  [CMS Collaboration],
  Phys.\ Rev.\ Lett.\  {\bf 111}, 101804 (2013)
  [arXiv:1307.5025 [hep-ex]].
\bibitem{Aaij:2012ac} 
  R.~Aaij {\it et al.}  [LHCb Collaboration],
  Phys.\ Rev.\ Lett.\  {\bf 108}, 231801 (2012)
  [arXiv:1203.4493 [hep-ex]].


\bibitem{cms-hdec}
  CMS Collaboration, CMS PAS HIG-14-002.

 \bibitem{Bechtle:2008jh}
  P.~Bechtle, O.~Brein, S.~Heinemeyer, G.~Weiglein and K.~E.~Williams,
  Comput.\ Phys.\ Commun.\  {\bf 181}, 138 (2010)
  [arXiv:0811.4169 [hep-ph]];
  Comput.\ Phys.\ Commun.\  {\bf 182}, 2605 (2011)
  [arXiv:1102.1898 [hep-ph]];
   PoS CHARGED {\bf 2012}, 024 (2012)
  [arXiv:1301.2345 [hep-ph]];
  Eur.\ Phys.\ J.\ C {\bf 74}, 2693 (2014)
  [arXiv:1311.0055 [hep-ph]].
\bibitem{Chatrchyan:2014tja} 
  S.~Chatrchyan {\it et al.}  [CMS Collaboration],
  arXiv:1404.1344 [hep-ex].
\bibitem{atlas-coup}
 ATLAS Collaboration, ATLAS-CONF-2014-010.
\bibitem{Aad:2014iia} 
  G.~Aad {\it et al.}  [ATLAS Collaboration],
  Phys.\ Rev.\ Lett.\  {\bf 112}, 201802 (2014)
  [arXiv:1402.3244 [hep-ex]].

\bibitem{Giardino:2013bma} 
  P.~P.~Giardino, K.~Kannike, I.~Masina, M.~Raidal and A.~Strumia,
  JHEP {\bf 1405}, 046 (2014)
  [arXiv:1303.3570 [hep-ph]].
  J.~Ellis and T.~You,
  JHEP {\bf 1306}, 103 (2013)
  [arXiv:1303.3879 [hep-ph]].

  
\bibitem{Belyaev:2012qa} 
  A.~Belyaev, N.~D.~Christensen and A.~Pukhov,
  Comput.\ Phys.\ Commun.\  {\bf 184}, 1729 (2013)
  [arXiv:1207.6082 [hep-ph]].
  
\bibitem{Staub:2013tta} 
  F.~Staub,
  Comput.\ Phys.\ Commun.\  {\bf 185}, 1773 (2014)
  [arXiv:1309.7223 [hep-ph]].
  
\bibitem{Belanger:2013oya} 
  G.~Belanger, F.~Boudjema, A.~Pukhov and A.~Semenov,
  Comput.\ Phys.\ Commun.\  {\bf 185}, 960 (2014)
  [arXiv:1305.0237 [hep-ph]].
  
\bibitem{Porod:2003um} 
  W.~Porod,
  Comput.\ Phys.\ Commun.\  {\bf 153}, 275 (2003)
  [hep-ph/0301101].
  
  

\bibitem{Barate:2003sz} 
  R.~Barate {\it et al.}  [LEP Working Group for Higgs boson searches and ALEPH and DELPHI and L3 and OPAL Collaborations],
  Phys.\ Lett.\ B {\bf 565}, 61 (2003)
  [hep-ex/0306033].
\bibitem{Aprile:2012zx} 
  E.~Aprile [XENON1T Collaboration],
  arXiv:1206.6288 [astro-ph.IM].
\bibitem{Daylan:2014rsa} 
  T.~Daylan, D.~P.~Finkbeiner, D.~Hooper, T.~Linden, S.~K.~N.~Portillo, N.~L.~Rodd and T.~R.~Slatyer,
  arXiv:1402.6703 [astro-ph.HE].
\bibitem{Aad:2014aba} 
  G.~Aad {\it et al.}  [ATLAS Collaboration],
  arXiv:1406.3827 [hep-ex].
\bibitem{Khachatryan:2014ira} 
  V.~Khachatryan {\it et al.}  [ CMS Collaboration],
  arXiv:1407.0558 [hep-ex].

 
\bibitem{atlasmu}
 ATLAS Collaboration, ATLAS-CONF-2014-009.
\bibitem{cmsz}
  S.~Chatrchyan {\it et al.}  [CMS Collaboration],
  Phys.\ Rev.\ D {\bf 89}, 092007 (2014)
  [arXiv:1312.5353 [hep-ex]].
\bibitem{cmsw}
  S.~Chatrchyan {\it et al.}  [CMS Collaboration],
  JHEP {\bf 1401}, 096 (2014)
  [arXiv:1312.1129 [hep-ex]].
\bibitem{cmsfer}
  S.~Chatrchyan {\it et al.}  [CMS Collaboration],
  Nature Phys.\  {\bf 10} (2014)
  [arXiv:1401.6527 [hep-ex]].
  

\end{thebibliography}
\end{document}